# Purcell–enhanced single photon source based on a deterministically placed WSe$_2$ monolayer quantum dot in a circular Bragg grating cavity

O. Iff[1], Q. Buchinger[1], M. Moczała-Dusanowska[1], M. Kamp[1], S. Betzold[1], S. Tongay[2], C. Antón-Solanas[3,*], S. Höfling[1], C. Schneider[1,3,*]

[1]*Technische Physik, Physikalisches Institut and Wilhelm Conrad Röntgen-Center for Complex Material Systems, Universität Würzburg, Am Hubland, D-97074 Würzburg, Germany*
[2]*School for Engineering of Matter, Transport, and Energy, Arizona State University, Tempe, Arizona 85287, USA*
[3]*Institute of Physics, University of Oldenburg, D-26129 Oldenburg, Germany*
*carlos.anton-solanas@uni-oldenburg.de; christian.schneider@uni-oldenburg.de

**We demonstrate a deterministic Purcell-enhanced single-photon source realized by integrating an atomically thin WSe$_2$ layer with a circular Bragg grating cavity. The cavity significantly enhances the photoluminescence from the atomically thin layer, and supports single-photon generation with $g^{(2)}(0)<0.25$. We observe a consistent increase of the spontaneous emission rate for WSe$_2$ emitters located in the center of the Bragg grating cavity. These WSe$_2$ emitters are self-aligned and deterministically coupled to such a broadband cavity, configuring a new generation of deterministic single-photon sources, characterized by their simple and low–cost production and intrinsic scalability.**

## INTRODUCTION

Advancing on the development of deterministic single-photon sources for quantum optical applications requires engineering new devices with reduced complexity in both production and operation. The benchmarking performance for deterministic quantum sources of light is brightness, single photon purity, and photon coherence (indistinguishability) [1]. At the moment, and following these criteria, the leading technology are self-assembled InAs quantum dots (QDs) which are integrated either in monolithic [2–4], or open cavities [5]. The success of a particular single-photon source technology depends not only on performance criteria but also on the affordability, simple manufacturing, and reproducibility of the device and its compatible integration with other pre-existing photonic technologies. In this sense, semiconductor QDs - while providing the current state-of-the-art performance as single photon sources - are arguably scalable [6,7] and constitute a resource-demanding technology, restricted to few specialized research groups around the globe. This imposes a significant barrier to ensure a broad, and fast impact of this solid-state-based single photon technology in a short time-scale, both for quantum applications, as well as for academic research.

Recently, a promising single-photon emitter platform emerged from two-dimensional (2D) materials. Seminal work has demonstrated that localized excitons in monolayers (MLs) of transition metal dichalcogenides (TMDCs) can deliver single photons under a continuous wave, and pulsed excitation [8–12]. TMDC QDs can be controllably created using strain [13,14], enabling ordered arrays of TMDC QDs [14,15]. In principle, such method allows the scalable and deterministic integration of TMDC QDs in photonic micro- and nano-structures, which may be used for improving the emitter performance as single-photon sources. Recently, the deterministic coupling of TMDC quantum emitters to plasmonic nanostructures was reported [16–19]. While the broadband nature of plasmonic approaches, in conjunction with the strong field enhancement makes this approach suitable for the integration of WSe$_2$ sheets, it remains a complicated task to fully suppress undesired Ohmic losses in plasmonic implementations. Open microcavities based on conventional Bragg mirrors have been utilized to demonstrate the weak coupling regime between a WSe$_2$ QD and a cavity mode [20]. However, this implementation, despite its flexibility in spectral tuning, presents challenges for applications that require large-scale TMDC QD integration or lab-independent user-friendliness.

An ideal cavity platform features a broad bandwidth operation with a large Purcell enhancement. This is accomplished when the cavity has simultaneously a moderate cavity quality factor (providing the large bandwidth) and a small mode volume (enhancing the resulting Purcell factor). Different types of photonic structures, which are compatible with the integration of atomically thin sheets, are natural towards this goal, including photonic crystal cavities [22] and cavities consisting of a circular disk surrounded by a set of concentric rings in the thin membrane. The latter is known in the literature as Circular Bragg grating cavities (CBGs). CBGs are especially interesting due to their simple design, accessing different emission wavelengths in the range of 600 nm to 1500 nm, with minimal design modifications. Besides, they can be designed to specifically maximize the electromagnetic cavity field at the cavity–air interface, where the TMDC QD is positioned [23].

In this work, we implement a hybrid CBGB composed of an Al$_{0.31}$Ga$_{0.69}$As membrane, whose cavity mode couples to a WSe$_2$ monolayer QD. These QDs are strain-defined in the monolayer by selectively grown nanopillars, centered in the CBGB. We demonstrate a spontaneous emission enhancement from the WSe$_2$ QD photoluminescence by more than a factor of ten. Moreover, we prove the single-photon character of this emission by performing a second-order correlation measurement, revealing a strongly suppressed multiphoton emission.

**DEVICE DESIGN AND FABRICATION**

Figure 1a) shows a schematic of our CBGB structure, starting from an epitaxially-grown 230 nm thick Al$_{0.31}$Ga$_{0.69}$As membrane. The membrane is bonded via the "flip-chip" method on 430 nm thick layer of SiO$_2$, on top of a 150 nm Au mirror, and a standard GaAs substrate. Details on the bonding procedure can be found in Ref. [24].

In the next steps, high-resolution electron-beam lithography and a subsequent lift-off process are used to prepare the pattern of the CBGBs. They are composed of five concentric rings, with a gap width of 100 nm and a period of 350 nm, including the 250 nm non-etched SiO$_2$ slab: a vertical-cut sketch of the CBGB, and its SEM image are shown in Figs. 1(a,b), respectively. The pattern is transferred to the semiconductor membrane using reactive ion etching with an Ar/Cl$_2$ plasma. To induce a QD "seed", we apply a local electron beam deposition method in a FIB to grow a 200 nm high SiO$_2$ nanopillar accurately in the center of the cavity, utilizing a precursor gas (Si(C$_2$H$_5$O)$_4$). This pillar serves to induce the WSe$_2$ QD via local strain formation (see Supplementary S1 for further details on the nanopillar processing).

For a quantitative assessment of the mode profile and the CBGB parameters, we perform numerical simulations using the finite-difference time-domain method. In Fig. 1(c) we plot the CBGB emission far field for an emitting dipole placed in the CBGB center and 10 nm above the membrane surface, emitting at a wavelength of 748 nm. The CBGB generates a strongly directed far field emission, optimal for collection with low numerical aperture optics, and hence constitute an efficient light-matter interface. The sideview of the absolute electric field in the membrane shows a strong confinement of the field in the innermost slab in Fig.1(d). Importantly, the thickness of the AlGaAs membrane is chosen so that it supports a second-order transverse-electric (TE$_2$) slab waveguide mode, with two vertical maxima, from which the bullseye cavity mode is formed (see Fig. 1(d) and Fig. S2). The longer evanescent tail of the TE$_2$ slab mode results in a cavity mode with a higher intensity above the AlGaAs slab, and therefore an improved coupling efficiency to emitters located at the surface of the structures. Cavities in which the fundamental (TE$_1$) slab mode is used to form a bullseye resonance offer lower coupling to emitters at the surface, and are better suited for emitters located at the slab center.

Due to the lower refractive index contrast from the membrane to the SiO$_2$ substrate, as compared to the air interface, the field preferentially extends into the substrate, however is reflected by the Au layer. The thickness of the SiO$_2$ is optimized to maximize the extraction efficiency out of the sample. In panel (e) we present the Purcell enhancement as a function of emission wavelength: a Purcell factor as large as 14 can be theoretically obtained. Apart of the promising mode at 746 nm, the spectrum in Fig 1e) reveals two weaker modes at 760 nm and 774 nm. The electromagnetic field distribution of those two modes can be found in Suppl. S2.

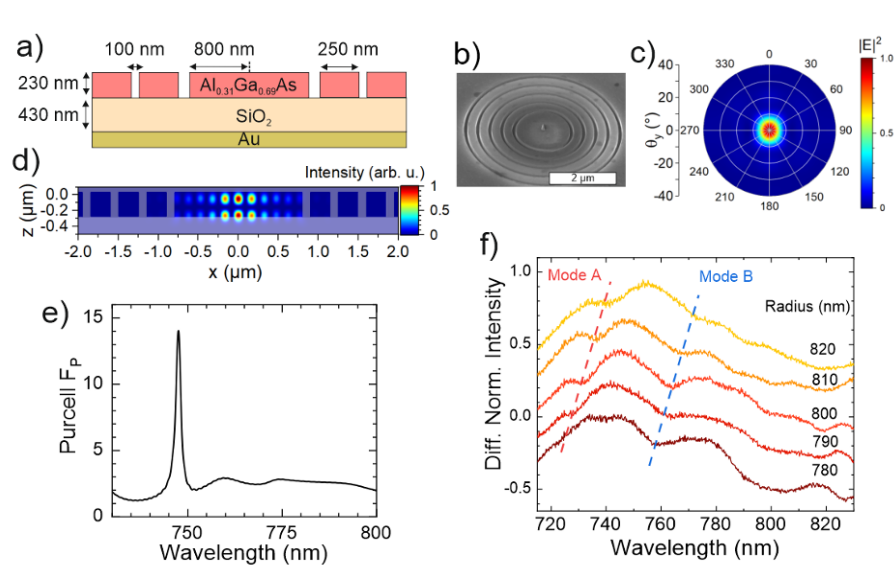

Fig 1. a) Schematic of the structure in a vertical cut. b) SEM image of a processed CBGB with a 200 nm pillar placed right in the centre, the white bar indicates 2 μm. c) FDTD far-field calculation of the electric field for a dipole placed in the centre for the structure, showing an emission with a narrow angular dispersion (<10°). d) FDTD simulation of the electric field in a vertical cut along the CBGB diameter. The photonic mode is mainly confined in the central disk. e) Simulation of the Purcell enhancement versus wavelength for the previously described CBGB. f) Reflectivity as a function of wavelength on the finalized sample varying the innermost radius (from 780 to 820 nm, see legend inside the panel). The spectral position of modes A and B are highlighted with dashed lines. This incremental variation reveals two modes in the range of 730 nm and 765 nm, which shift red as a function of increasing radius.

The CBGB mode structure is clearly visualized via reflectivity measurements, where the photonic modes show reflectivity dips. In Fig. 1(f), we depict the results of a study comprising several CBGBs with varying nominal inner disk radius from 780 nm to 820 nm while the other parameters stay constant. We plot the corresponding white-light micro-reflectivity spectra after normalizing to the reflection from a silver surface. The spectral position of the cavity modes A and B are highlighted with dashed lines in Fig. 1(f), shifting as a function of the inner disc diameter. The mode A (B) displays a bandwidth of 19 nm (27 nm) and a quality factor of ~38 (28), spectrally overlapping with the typical energy emission of WSe$_2$ QDs (720-780 nm). By comparing these spectra with the FDTD simulation in Fig 1(e) and Fig S2, we assign mode A to the resonance that exhibits a larger Purcell enhancement.

The single photon sources are finalized by the deposition of a monolayer of WSe$_2$. These monolayers are exfoliated from a bulk crystal and transferred with a polymer (PDMS) stamp onto the device [25]. To enhance the adhesion of the monolayer to the CBGB, the stamping procedure was conducted at a temperature of 150 °C.

**RESULTS AND DISCUSSION**

To facilitate a global characterization of our structures, we selected particularly large monolayers on our PDMS stamp, with sizes exceeding 50 μm x 50 μm. Using flakes of that size, a single monolayer covers various CBGBs. In Fig. 2a) we present an optical microscope image of such a monolayer, which covers 4 CBGBs (the yellow line indicates the borders of the monolayer as a guide to the eye). In fig 2b), the close-up SEM image verifies that the nanopillar in the cavity center produces a nanometric "tent", locally straining the monolayer, and so, producing single-excitonic emitters (WSe$_2$ QDs). We reveal these single-excitonic sharp energy transitions by studying their micro-photoluminescence at cryogenic temperatures (4K) in a helium flow cryostat. The cryostat is mounted on a set of closed-loop X-Y translation stages, which allows performing two-dimensional scans with a spatial accuracy of ~100 nm.

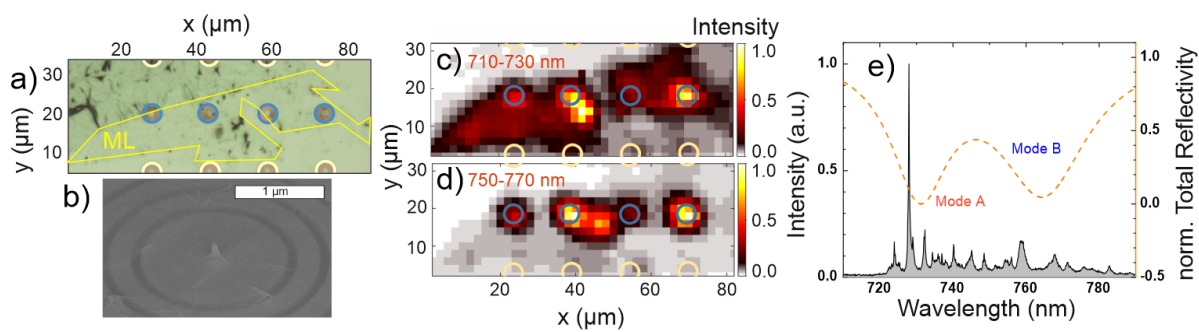

Fig. 2. a) Microscope image the transferred monolayer (its limits are highlighted with a yellow line as a guide to the eye), covering several CBGBs with an inner radius of 800 nm (these cavities are highlighted with open blue circles). b) SEM image of a transferred WSe$_2$ monolayer placed on the centre of a CBGB. The pillar is fully covered by the monolayer, some monolayer wrinkles are visible. c/d) Photoluminescence intensity scans of the monolayer shown in panel a), integrating the energy mission in the spectral range 710 nm to 730nm/ 750 nm to 770 nm, respectively. The emission is enhanced in the centre of each CBGB. e) Spectrum of one of the CBGBs shown in b). The simulated CBGB reflectivity is shown on the right axis, observing the spectral location of mode A and B. The spectral position of the dominant line at 728 nm coincides with the position of mode A.

We record hyperspectral maps of the sample by scanning the sample position in steps of 2 μm while exciting the structure non-resonantly with a 532 nm laser (12 W/cm²) to capture the inter-dependence between luminescence intensity, sample position, and spectral window of interest. Figures 2c,d) show a photoluminescence map resulting from the scan of the 4 single photon sources; these scans integrate the energy emission from 710 nm to 730 nm and

750 nm to 770 nm, respectively. The position of the covered CBGBs are schematically highlighted with blue circles.

In Fig. 2c), we notice that in this spectral range 710 nm to 730 nm, apart from the sharp QD emission features, the WSe$_2$ luminescence is characterized by the free exciton and the free trion band (in the Supplementary information Fig. S3 we show a spectrum outside of the CBGB for the sake of comparison). As a result, significant photoluminescence is observed over the entire monolayer area. A slight enhancement of the luminescence at the positions of the CBGBs occurs in this spectral range. In Fig. 2d) on the contrary, the background luminescence from the WSe$_2$ monolayer is strongly reduced in the 750 nm to 770 nm range: the photoluminescence intensity almost exclusively arises from the cavity positions. This clearly demonstrates the capability of the CBGBs to efficiently redirect and enhance the single-photon emission in our desired collection mode.

A typical photoluminescence spectrum from one of the CBGBs is shown in Fig. 2e). The nanopillar yields the emergence of a variety of significant sharp emission features, which occur in the spectral region between 720 nm and 800 nm. We include in this graph the fitted reflectivity spectrum of the corresponding CBGB, where the spectral alignment of mode "A" with the enhanced emission intensity at 730 nm suggests an acceleration of the spontaneous emission of this emitter via the Purcell effect. This effect is also reflected in a comparison of the power dependent photoluminescence of several emitters from different positions on the flake, indicating a higher saturation brightness for emitters on the CBGB (see Supplement Fig. S4).

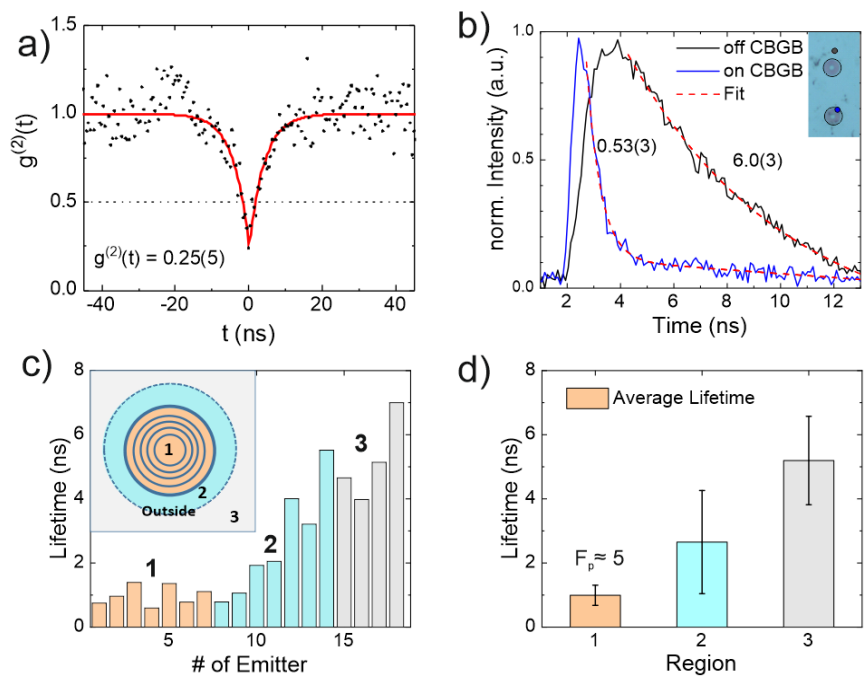

Fig. 3. a) Second-order correlation measurement from a selected spectral line (shown in the supplementary) in a CBGB. b) Normalised lifetime measurement on (blue) and off (grey) the CBGB. c) Statistic of 18 individual emitters located in the monolayer, sorted by their position: 1) inside the inner disk 2) in the outer limit of the CBGB border, and 3) outside the CBGB device. The lifetimes in the region 1 (orange) are close to the detector resolution (350 ps), in contrast to emitters outside of the device (grey bars). d) Average lifetime of emitters from the same region, including the corresponding estimated Purcell enhancement relative to region 3. Error bars correspond to one standard deviation.

To verify the single-photon character of the investigated CBGBs, a second-order correlation experiment is performed in one of the intense, sharp emission lines (see Supp. S5, where the corresponding filtered spectral line is highlighted with a gray band). Figure 3a) shows the normalized second-order correlation function $g^{(2)}(\tau)$ histogram under continuous-wave excitation. The strong reduction of the double-coincidence events at zero delay proves the population suppression of multi-photon Fock states in the emission, and a predominant generation of single photons. The mono-exponential fit in Fig. 3a) shows a $g^{(2)}(0)=0.25(5)$, which puts our emitter clearly in the class of a single photon source.

We now study the spontaneous emission acceleration experienced by cavity-coupled $WSe_2$ emitters. As the cavity bandwidth is too broad to study the same emitter on- and off-resonance by applying an external tuning knob (such as temperature or mechanical stress), we perform comparative measurements of $WSe_2$ emitters emerging spatially inside and outside the cavities. The lifetime measurements are performed with a frequency-doubled Titanium Sapphire laser providing pulses of 2 ps at a wavelength of 448 nm with a repetition rate of 75 MHz. The single photon detectors are silicon APDs presenting a time resolution of 350 ps.

Figure 3b) shows the normalized decay curves of two emitters located inside (blue curve) and outside (red curve) the CBGB. Each data set is fitted with a mono-exponential decay curve, yielding a characteristic decay time of 0.5(3) ns and 6.9(4) ns, respectively. The difference in decay time suggests an accelerated spontaneous emission rate, as a result, we predict a Purcell enhancement of ~10 for emitters efficiently coupled to the CBGB.

For a more quantitative analysis of this observation, in Fig. 3c) we subdivide the CBGB in three different concentric regions, numbered from 1 to 3, and we analyze the lifetime of 18 emitters distributed in these regions. Notably, the acceleration of the emitter dynamics systematically occurs throughout our selected structures in region 1, which is restricted to the area inside of the CBGB and where a good spectral overlap of the emitter positions and the cavity modes are found (see Suppl. Fig. S6). Region 2 represents the outer rim, where a determination of the exact emitter position within the laser excitation spot (FWHM ≈ 3 µm) is difficult. Emitters in region 3 are clearly outside of the CBGB. The resulting histogram of the average lifetime as a function of the region, see Fig. 3d), reveals that the emitters located in the region 1 consistently experience a short lifetime of 0.99±0.31 ns, whereas the average lifetime in region 2 and 3 displays a lifetime of 2.7±1.6 ns and 5.2±1.4 ns, respectively (uncertainties correspond to one standard deviation among the measured emitters in each region). Assuming that a $WSe_2$ emitter on the plane AlGaAs surface (uncoupled to a CBGB) does not experience any modification of its spontaneous emission, this analysis determines a Purcell factor of approximately up to 5 in these CBGB structures.

It is worth noting that, in contrast with previous reports on lifetime acceleration of $WSe_2$ emitters using metallic nanostructures, the emergence of non-radiative losses due to the presence of dielectric nanopillars is less likely to play a critical role in the observed phenomena: $WSe_2$ quantum emitters induced by polymer- or silicon nanopillars without the presence of CBGB structures have consistently been reported to feature spontaneous emission lifetimes in the range of 3 to 8 ns, putting them consistently in category 3 of our emitter assignment [14,15]. This strongly suggests that observed phenomenon of lifetime reduction in our work can be attributed to the amplification of spontaneous emission in the weak coupling regime between the $WSe_2$ QD and the CBGB.

# CONCLUSION

To conclude, we have implemented a deterministically coupled, strain engineered single-photon source by integrating a $WSe_2$ QD in a circular Bragg grating Bullseye cavity. Our cavities provide a strong enhancement of spontaneous emission from $WSe_2$ quantum emitters, and thus, our studies systematically demonstrate the first prototype of a monolithically coupled, cavity-enhanced single photon source, based on a two-dimensional material. We believe that the performance of our devices can be further improved by utilizing resonant excitation and combining the cavity with other external knobs such as mechanic strain [26] or electric control of the QD environment [11].

# ACKNOWLEDGMENTS

We gratefully acknowledge support by the State of Bavaria. This work has been supported by European Research Council within the Project unLiMIt-2D (Project No. 679288) and by the German Research Association within the project PR1749 1-1. C.S. acknowledges fruitful discussions and creative input by R. Trotta and A. Predojevic as well as support by P. Pertsch and Prof. B. Hecht, University of Würzburg, for assisting within the dry stamping procedure. C.S. acknowledges the help of S. Kuhn, who carried out the lithography process, and M. Emmerling, who took the SEM images. S.T acknowledges support from DOE-SC0020653, NSF CMMI 1933214, NSF mid-scale 1935994, NSF 1904716, NSF DMR 1552220 and DMR 1955889.

**SUPPLEMENTARY MATERIAL**

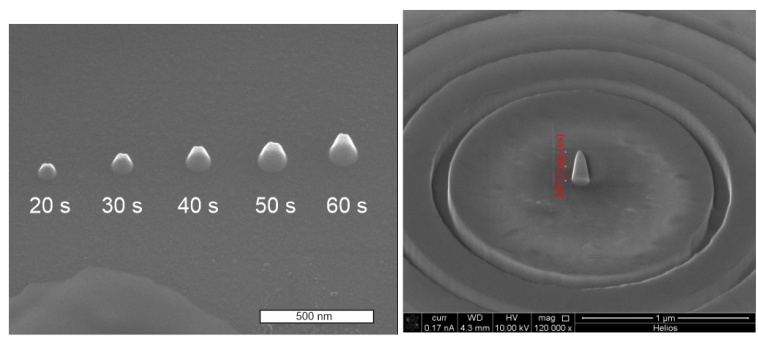

Fig. S1. (left) SEM image of the systematic calibration on the deposited Si nanopillar. From left to right, the five nanopillars present the following base and height dimensions (±5 nm): [62, 80, 82, 98, 120] nm and [83, 97, 160, 168, 257] nm, respectively. (right) Similar to Fig. 1(b) of the main text, SEM image of a finalized CBGB, ready for a $WSe_2$ monolayer transfer, the corresponding height of the nanopillar is indicated in the figure > 200 nm.

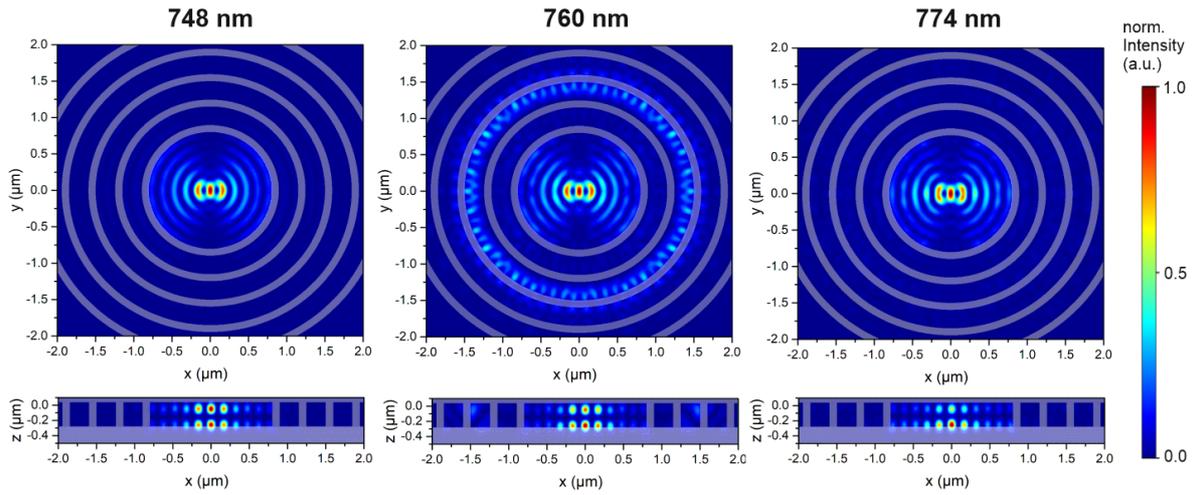

Fig. S2. FDTD comparison between the 3 modes visible in the Purcell calculation in Fig. 1e) at 748/760/774 nm. Both the top view (first row) and the side view (second row) reveal a mainly mode confinement to the inner disk. Each graph is normalized to [0,1].

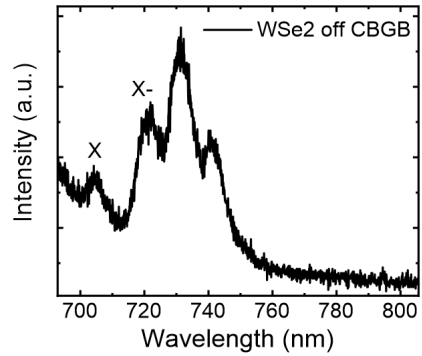

Fig. S3. A representative PL spectrum of the WSe2 monolayer flake off the CBGB cavity. The typical WSe2 features at 4K are visible, with the exciton (X) and the trion (X-) being assigned based on their spectral positions.

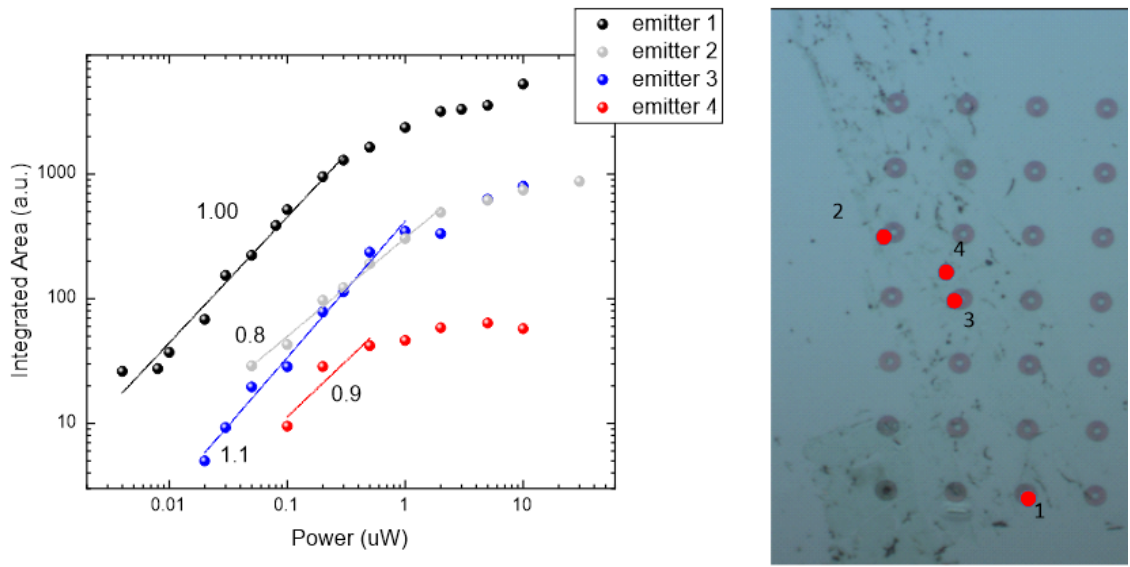

Fig. S4. (left) Pump power dependence of the photoluminescence intensity from four different emitters with their corresponding fitted slopes. (right) Microscope image of the corresponding emitters 1-4 used for the left

panel. The devices 1-3 show similar trends with similar pump power saturation values; in contrast, emitter 4 shows a rapid saturation.

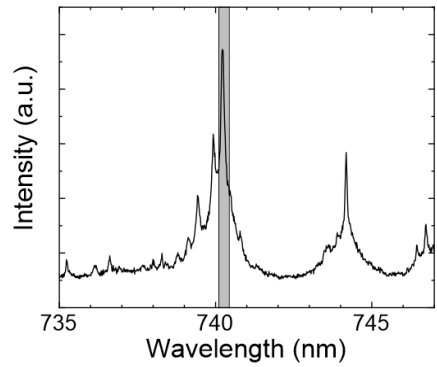

Fig. S5. Spectrum of the emitter used for the second-order correlation measure reported in Fig. 3(a) of the main text. The spectrally post-selected line is highlighted in a grey band.

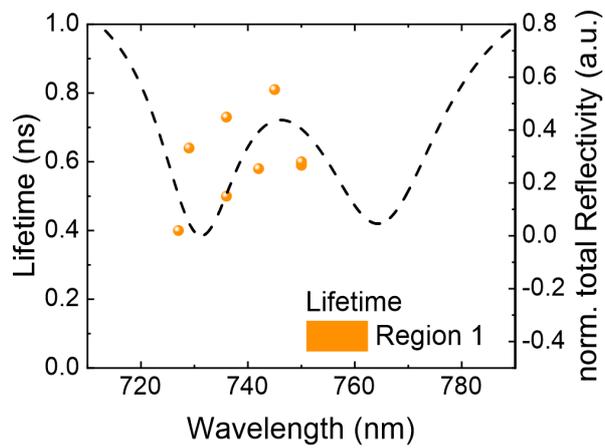

Fig. S6. Detuning of all region 1 emitters measured in Fig. 3c) overlayed with the cavity modes found in the reflectivity measurements. Error of the spectral position of each emitter is about ±0.5 nm due to jitter.